# Spatial Segregation of Substitutional B Atoms in Graphene Patterned by the Moiré Superlattice on Ir(111)

*Marc G. Cuxart,[‡][*] Daniele Perilli,[‡] Sena Tömekce, Joel Deyerling, Felix Haag, Matthias Muntwiler, Francesco Allegretti, Cristiana Di Valentin,[*] and Willi Auwärter[*]*

Fabrication of ordered structures at the nanoscale limit poses a cornerstone challenge for modern technologies. In this work we show how naturally occurring moiré patterns in Ir(111)-supported graphene template the formation of 2D ordered arrays of substitutional boron species. A complementary experimental and theoretical approach provides a comprehensive description of the boron species distribution, bonding configurations, interfacial interaction with Ir(111) and the impact on graphene's electronic structure. Atomically-resolved scanning tunnelling microscopy images and density functional theory calculations reveal that boron preferably forms small aggregates of substitutional defects in geometrically low regions of the moiré superlattice of graphene, by inducing local bending of graphene towards the underlying Ir(111). Surprisingly, calculations reveal that the incorporation of electron deficient boron does not lead to an enhanced p-type doping, as the local rippling of the graphene layer prompts electron uptake from the iridium substrate that compensates the initial electron loss due to substitutional boron. Scanning tunnelling spectroscopy and angle-resolved photoemission spectroscopy measurements corroborate that the arrays of boron species do not modify the electronic structure of graphene near the Fermi level, hence preserving the slight p-type doping induced by Ir(111).

KEYWORDS: Boron doping, graphene, nano-fabrication, moiré superlattice, CVD, XPS, STM, STS, ARPES, DFT.

## 1. Introduction

Graphene grown on metallic substrates commonly reveals moiré interference patterns due to the mismatch between the constituting atomic lattices. The graphene carbon atoms occupy inequivalent positions on the underlying metal, inducing interaction of distinct strengths or even nature, concomitantly promoting a periodic modulation of graphene's corrugation and electron density [1]. As a result, moiré superlattices constitute ordered nanoscale structures, otherwise exceptionally challenging to achieve at such small scales [2], which possess intriguing quantum properties [3,4], and catalytic or templating functionalities. In this context, extensive research efforts have been invested on site-selective adsorption of atomic clusters [5–7] or organic molecules on graphene [8–11] whereas far less attention has been devoted to templating the incorporation of heteroatoms into the epitaxial monolayers.

Using moiré patterns to achieve control over the distribution of heteroatoms embedded in graphene shall constitute a way to fabricate very robust ordered nanoscale structures [12], and to modify graphene's properties permanently [13]. Boron and nitrogen are the most commonly explored substitutional species because of their proximity to carbon in the periodic table, the similar size, and interest as acceptor or donor centers, respectively [14]. Diverse experimental methods have been proposed based on simultaneous graphene synthesis and integration of heteroatoms [15–20], or post-synthetic incorporation of such species [21–26]. While in the vast majority of cases heteroatoms distribute randomly within graphene, the use of strongly corrugated graphene supported on Ru(0001) has shown promising results towards site-selective pinning of substitutional nitrogen species at low densities [27].

Here, we show how the strongly corrugated moiré patterns present in graphene on Ir(111) template the assembly of substitutional boron species, thus creating a 2D array of small boron aggregates within graphene. . Through a comprehensive combined experimental and theoretical study, based on a single-atom and surface-averaged characterization and density functional theory calculations, a complete description and rationalization of heteroatom distribution, B-C bonding configuration, interfacial interaction, and impact on graphene's electronic structure is provided. Interestingly, we find that the site-selective incorporation of boron species induces local bending of the graphene layer towards the Ir(111) substrate, which reverses the charge transfer at the graphene-Ir(111) interface and compensates for the hole doping induced by substitutional boron. Such a unique doping mechanism results in a 2D periodic modulation of enhanced geometric rippling and charge accumulation in graphene, highly appealing for site-selective manipulation of molecular states [28], and applications such as compact photo-catalytic devices and gas detectors [2,29].

## 2. Experimental and theoretical methods

The experiments were performed in two different ultra-high vacuum setups, one at the Technical University of Munich (TUM, Garching, Germany) dedicated to the low-temperature scanning tunnelling microscopy (STM) measurements, and the other one at the PEARL beamline (Swiss Light Source, Villigen, Switzerland) [30] devoted to the X-ray photoelectron spectroscopy (XPS) and angle-resolved photoemission spectroscopy (ARPES) experiments. Both chambers were equipped with a low energy electron diffraction (LEED) apparatus, which provides a link between the experiments. Samples were reproducibly grown *in-situ* in each chamber.

### 2.1. Sample preparation

The Ir(111) single crystal (6 mm diameter surface, Mateck GmbH) was prepared by repeated cycles of sputtering ($Ar^+$ ions at an energy of 1 keV), followed by two-step annealing at 1173 K under $O_2$ atmosphere ($3\times10^{-8}$ mbar) and at 1233 K subsequently. The boron-doped graphene samples were grown by dosing 5.2 L of borane tetrahydrofuran (BTHF, $BH_3OC_4H_8$) on Ir(111), pre-heated at temperatures between 1073 - 1233 K. The resulting graphene monolayer presented a manifold of rotational domains described in the text, with typical lateral sizes of tens of nm, whose occurrence followed the same temperature dependency as when grown from ethylene precursor [31]. No temperature-dependent variation in the distribution and density of boron species was observed. Un-doped graphene was grown by dosing 5.4 L of ethylene on Ir(111) pre-heated at 1123 K. Commercially available BTHF (1.0 M in THF solution, Sigma-Aldrich) and ethylene (99.95%, CAN-Gas) were used.

### 2.2. STM and STS characterization

STM and scanning tunneling spectroscopy (STS) data were acquired using a CreaTec STM operating at 6 K and $P < 2 \times 10^{-10}$ mbar. STM images were taken in constant current mode, at the bias voltage stated in the figure captions, and processed using the WSxM software [32]. d$I$/d$V$ spectra were acquired in constant height mode with a lock-in technique at 961 Hz and 5 mV peak-to-peak modulation voltage.

### 2.3. XPS and ARPES characterization

Photoelectron intensity was monitored by a Scienta EW4000 hemispherical electron energy analyzer with two-dimensional detection, with the sample held at 77 K and operating at $P < 2 \times$ $10^{-10}$ mbar base pressure. Spectra were taken in normal emission (unless stated otherwise in the figure captions) with photon energies of 680 eV (O 1s), 390 eV (C 1s and B 1s) and 60 eV (valence band). Binding energies were calibrated using the Fermi edge subsequently measured at the same conditions. Core-level and valence-band spectra were measured at 10 and 20 eV pass energy, respectively. Fits to the core-level spectra were performed using the XPST macro for IGOR (M. Schmid, Philipps University Marburg), using Voigt-like functions on a Shirley background. Fitting parameters are summarized in **Table S2**.

### 2.4. Computational details

DFT calculations have been performed using the plane-wave-based Quantum ESPRESSO package (QE) [33,34]. The ultrasoft pseudopotentials have been adopted to describe the electron-ion interactions with C (2s, 2p), B (2s, 2p), and Ir (4s, 3d) treated as valence electrons [35]. Energy cutoffs of 43 and 361 Ry (for kinetic energy and charge density expansion, respectively) have been adopted for all calculations. The convergence criterion of 0.026 eV/Å for the forces has been used during geometry optimization, and the convergence criterion for the total energy has been set at $10^{-6}$ Ry. To properly take into account weak interactions, the van der Waals density functional vdW-DF2$^{C09x}$ has been used [36,37], as implemented in the QE code.

The Ir(111) surface was modeled by a three-layer slab model with the two bottom layers fixed to the bulk positions during the geometry relaxation to mimic a semi-infinite solid.

For the simulation of the graphene/Ir(111) interface, we used the most commonly experimentally observed domain (R0), which is the (10 × 10) graphene cell on a (9 × 9) Ir(111) surface. A Monkhorst-Pack [38] k-Points mesh of 2 × 2 × 1 and 6 × 6 × 1 was used for the geometry relaxation and density of states (DOS) evaluation, respectively. To avoid interactions between adjacent periodic images, a vacuum space of 17 Å (along the z-direction) has been included in the supercell model. Dipole correction was employed to check field effects but found to be negligible and therefore it was not included.

STM simulations were performed using the Tersoff-Hamann approach [39], according to which the tunneling current is proportional to the energy-integrated Local Density of States (ILDOS). STM images were rendered with Gwyddion software [40]. Constant-current and voltage values for the STM simulations have been chosen to match the experimental values.

The XPS spectra of boron-doped graphene on Ir have been calculated through the ΔSCF approach [41], which includes a pseudopotential generated with a full core hole for each ionized inequivalent carbon atom and allows the calculation of relative changes of binding energies, called core-level shifts (CLSs). In order to compare the CLSs of different carbon atoms bounded to boron, we considered a carbon atom in the TOP region as a reference, which is far away from the dopants and therefore is not perturbed by the substitutional boron atoms in graphene, as discussed in the main text.

## 3. Results and discussion

The boron-doped graphene monolayer was grown on the (111) surface of an iridium single crystal, which was kept at 1073 K during exposure to 5.2 Langmuir (L) of BTHF gas under ultra-high vacuum conditions. BTHF was chosen as molecular precursor as it contains both carbon and boron elements, hence enabling the introduction of boron substitutional atoms simultaneously to the growth process of graphene. Ir(111) was chosen as support because of the strong periodic corrugation that it naturally induces to graphene, which constitutes a promising platform where the segregation of boron substitutional atoms can be templated.

The resulting material was first characterized *in situ* by core level XPS in order to elucidate its elemental composition. The spectrum in **Figure 1**a shows an intense C 1s peak featuring a main component ($C_1$) centered at a binding energy of 284.09 eV, which is characteristic for graphene/Ir(111), as confirmed on a reference sample grown by CVD using ethylene [42] (see **Figure S1** in the Supplementary Materials). Thus, the novel BTHF precursor also affords the formation of monolayer graphene on Ir(111). In comparison to the reference graphene, the C 1s peak presented in **Figure 1**a exhibits a broader signature, which can be modelled by considering two additional contributions ($C_2$ and $C_3$) emerging due to the presence of multiple rotational domains concomitant with an increased density of domain boundaries [43]. Their presence is confirmed by real-space visualization provided by atomically-resolved STM (**Figure S4**), and by the emergence of a ring-like feature in the corresponding LEED pattern, (**Figure 2**e). In contrast, the reference graphene sample presents sharp spots aligned with the Ir(1 × 1) pattern (inset in **Figure S1**), consequence of the presence of a single rotational domain [44]. In addition to $C_1$, $C_2$ and $C_3$, a tiny component ($C_4$) is attributed to carbon bonded to boron, as it appears -1.59 eV below $C_1$, in line with the lower binding energies predicted by the core level shift (CLS) calculations presented for boron species bonding to carbon via $sp^2$ -hybrid orbitals (**Figure S2**) [19,45,46].

The presence of boron is confirmed by the B 1s core level observed in **Figure 1**b that accounts for a concentration of ∼ 2.5 %. It presents a slight asymmetry that can be modelled by two components, $B_1$ and $B_2$, positioned at 187.77 and 188.63 eV, respectively. The main $B_1$ component appears at the characteristic binding energy of substitutional boron atoms forming $sp^2$ bonds with carbon, as reported for boron-doped graphene supported on diverse substrates [19,45,47]. The minority $B_2$ component emerging at 188.63 eV can be attributed to boron bond to boron in the presence of an Ir(111) substrate [48]. Its small intensity compared to that of $B_1$ is consistent with the low occurrence of boron directly bonded to neighboring boron observed by atomically resolved STM, as discussed below based on **Figure 5**. The Ir $4f_{7/2}$ signal shown in **Figure 1**d is modelled by a main component that accounts for the Ir atoms in the bulk ($Ir_1$), and two additional components that originate from the outermost Ir atoms weakly ($Ir_2$) and strongly ($Ir_3$) interacting with the overlayer [43]. The possibility that the graphene layer is oxidized due to the presence of oxygen in the precursor molecule is ruled out due to the absence of any O 1s signal (**Figure 1**c), and of a carbon – oxygen component that would be expected at between +1.5 and +2.5 eV above C1 (**Figure 1**a).

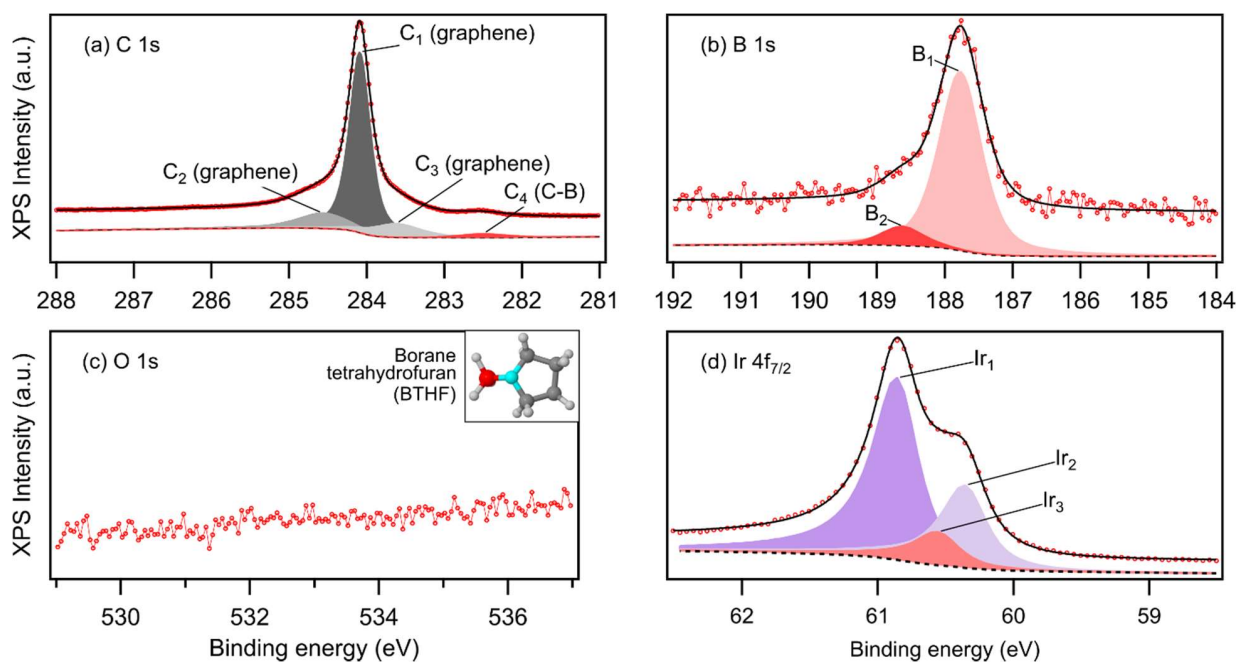

**Figure 1. Chemical characterization of boron doped-graphene on Ir(111).** High-resolution XPS of (a) carbon, (b) boron, (c) oxygen 1s and (d) iridium $4f_{7/2}$ core levels measured on boron-doped graphene on Ir(111). Experimental raw data are represented by red lines and circles, fits of the indicated components by colored areas and the sum of the fits by solid black lines. Raw data and sum of the fits are vertically offset for a better visualization. Inset in (c) contains a ball-and-stick model of the molecular precursor BTHF (carbon is depicted in dark grey, hydrogen in light grey, oxygen in cyan and boron in red).

Next, real-space imaging of the boron-doped graphene on Ir(111) with atomic resolution was performed by low-temperature STM. In the topographic image of **Figure 2**a, the graphene layer is readily identifiable due to its characteristic honeycomb structure (depicted in yellow) and the hexagonal moiré superlattice (in black) [49], which is a consequence of the lattice mismatch (∼9.3%) and relative azimuthal angle between the graphene lattice and the underlying Ir(111) surface. On numerous metal substrates, the moiré superlattice is accompanied by a strong corrugation of the graphene [50], thus defining regions in which graphene is adsorbed closer or further away from the substrate (here referred as "low" or "high" areas, respectively). In the present case, several graphene rotational domains exhibiting distinct moiré superlattices were identified by STM (see **Figure S4**), and are reflected in the ring-like feature around the spots of the Ir(111) substrate in the LEED patterns in **Figure 2**d,e. Specifically, R0, R17 and R18 domains (where the number denotes the rotation angle) with large moiré periodicity (24.3 ± 1.2, 23.0 ± 1.6 and 24.3 ± 1.2 Å respectively) and high degree of corrugation (∼ 20 pm), and R14 and R19 domains with smaller periodicity (10.9 ± 0.5 and 7.2 ± 0.4 Å) and corrugation (∼ 5 pm), as calculated by Meng *et al.*, [44] were observed. The rotational domain shown in **Figure 2**a,b can be further identified as R17 based on the angle (16.7 ± 0.3°) and ratio (0.11 ± 0.01 versus 0.112) between $q_{graphene}$ and $q_{moiré}$ extracted from the fast Fourier transform (FFT) **Figure 2**c. Here, we will focus on the highly corrugated domains R0, R17 and R18, and treat them indiscriminately because of their very similar moiré superlattice, and equal distribution and appearance of boron species (see **Figure S4**). The observation of such unique moiré superlattices, which are only visible by STM when monolayer graphene is in direct contact to Ir(111), demonstrates the single-atom thickness of the studied graphene.

Voloshina *et al.* [51] showed that the nature of the moiré corrugation of R0 is of both geometric and electronic origin, with geometrically high regions appearing as protrusions when imaged at bias voltages near the Fermi level by STM, and as depressions at higher voltages. This bias-dependent contrast inversion is also observed in the other highly corrugated domains, as shown in **Figure 2**a,b, where geometrically elevated regions are highlighted by black dashed circles. In addition to the moiré superlattice, the graphene layer presents multiple defects that appear as tiny "dents" of variable geometry at low bias voltage and broader depressions at higher bias voltage (highlighted by red dashed circles in **Figure 2**a and b, respectively). They will be identified as small aggregates of boron substitutional atoms (see below). These imperfections appear to have the tendency to avoid the geometrically elevated regions, that is they occupy intermediary regions within the moiré unit cell in which graphene is adsorbed closer to the supporting Ir(111). As a result, the defects assemble in a quasi-hexagonal array that mimics the periodicity and orientation of the moiré superlattice. The very same phenomenon is also observed in the highly corrugated domains R0, R17 and R18, but not in flat R14 and R19 areas, which appear completely defect-free (see **Figure S4**).

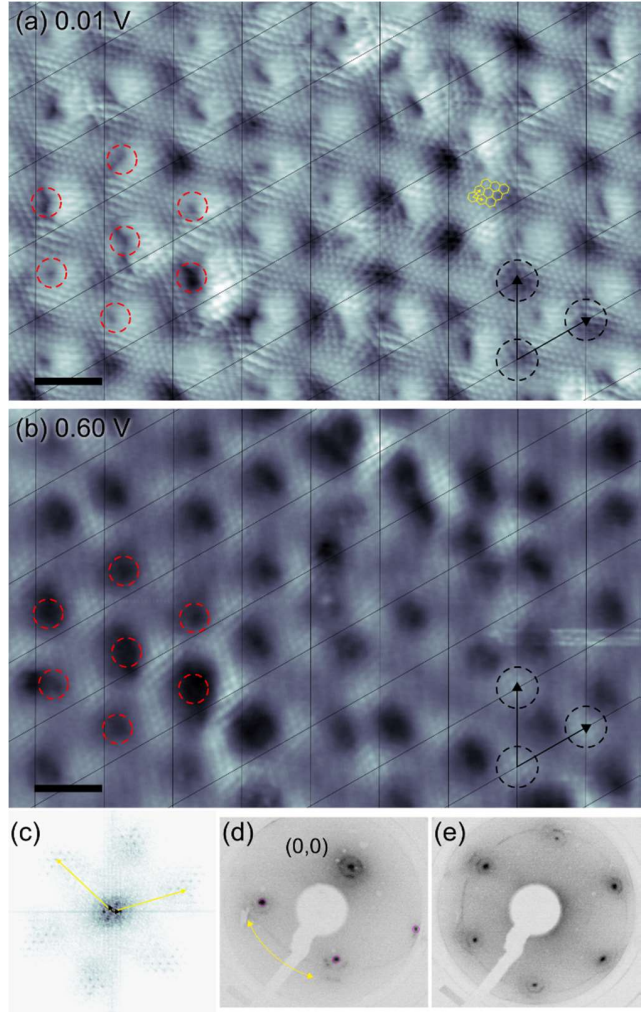

**Figure 2. Assembly of boron substitutional species in Ir(111)-supported graphene.** STM images of boron-doped graphene on Ir(111) imaged at (a) $V_{bias}$ = 0.01 V and (b) 0.60 V (scale bars: 2nm). Brighter/darker contrast indicates higher/lower topographic apparent height. Graphene honeycomb network is highlighted in yellow, moiré superlattice in black and boron defects in red. (c) FFT pattern calculated corresponding to the STM image in (a). $q_{graphene}$ and $q_{moiré}$ are indicated by yellow and black vectors. LEED patterns measured in the same sample at (d) 39 eV (e) 78 eV. Ir(111) (1×1) spots are marked by pink circles and ring-like feature from a manifold of graphene rotational domains by yellow arrows in (d).

In order to clarify the peculiar behavior of the corrugated rotational domains, we performed a wide set of density functional theory (DFT) calculations to get insight into the structure and relative stability of different possible configurations of substitutional boron ($B_C$) in the R0 domain of graphene/Ir(111). $B_C$ species are the most plausible candidates for the observed defects, since this element is present in the BTHF precursor and, unlike oxygen, was detected by XPS (**Figure 1**b). **Table 1** reports the relative energies obtained for single and multiple $B_C$ species occupying non-equivalent graphene sites (top, fcc, and hcp) within the super-cell model used, either in the geometrically high (*i.e.*, TOP in **Figure 3**a) or low (*i.e.*, FCC and HCP in **Figure 3**a) regions. As a first result, we note that geometrically high regions (**Figure 3**b) appear energetically less favorable than geometrically low ones (**Figure 3**c) for hosting individual $B_C$ species, forcing these to distribute in a specifically ordered pattern, according to the graphene moiré superlattice. Secondly, the aggregation of more than one $B_C$ in geometrically low regions (2 $B_C$ in FCC-hcp) is largely preferred, by more than 2 eV, with respect to a more spread distribution involving both geometrically low and high regions (1 $B_C$ in FCC-hcp + 1 $B_C$ in TOP-fcc) and has a negligible energetic toll, which should allow the formation of small $B_C$ aggregates (**Table 1**). The tendency for aggregation of the B atoms in the FCC region of the moiré can be further proved by comparing 3 $B_C$ in FCC-hcp with other configurations where one or two B atoms are moved to random positions far from the TOP, FCC and HPC regions (named "2 $B_C$ in FCC-hcp + 1 $B_C$ random" and "1 $B_C$ in FCC-hcp + 2 $B_C$ random", respectively). These configurations are higher in energy by +0.46 and +0.26 eV, respectively, as reported in **Table 1**.

**Table 1. Relative energies (in eV) of different $B_C$ configurations in graphene supported on Ir(111).** Energies are to be compared only for systems with the same type and number of atoms. (*i.e.*, same value of #$B_C$). A schematic representation of the different $B_C$ positions reported in the table is shown in **Figure S3**.

| #$B_C$ | Height | $B_C$ position | ΔE (eV) |
|---|---|---|---|
| | | | |

| 1 | Low | FCC-hcp | +0.29 |
| 1 | Low | HCP-fcc | 0.00 |
| 1 | High | TOP-fcc | +2.21 |
| 2 | Low | FCC-hcp | +0.09 |
| 2 | Low | FCC-hcp + HCP-fcc | 0.00 |
| 2 | High | TOP-fcc | +4.38 |
| 3 | Low | FCC-hcp | 0.00 |
| 3 | Low+Intermediate | 2 FCC-hcp + 1 random | +0.46 |
| 3 | Low+Intermediate | 1 FCC-hcp + 2 random | +0.26 |
| 3 | High | TOP-fcc | Not found |

It is also worth noting how $B_C$ species in the geometrically low regions (FCC and HCP), unlike in the high ones (TOP), cause significant geometric distortion of the graphene sheet, reducing the shortest vertical separation between the carbon layer and the Ir surface from ~ 3 to ~ 2 Å. This observation hints to an enhancement of the interaction between graphene and Ir(111) in the presence of substitutional boron atoms, as will be discussed below.

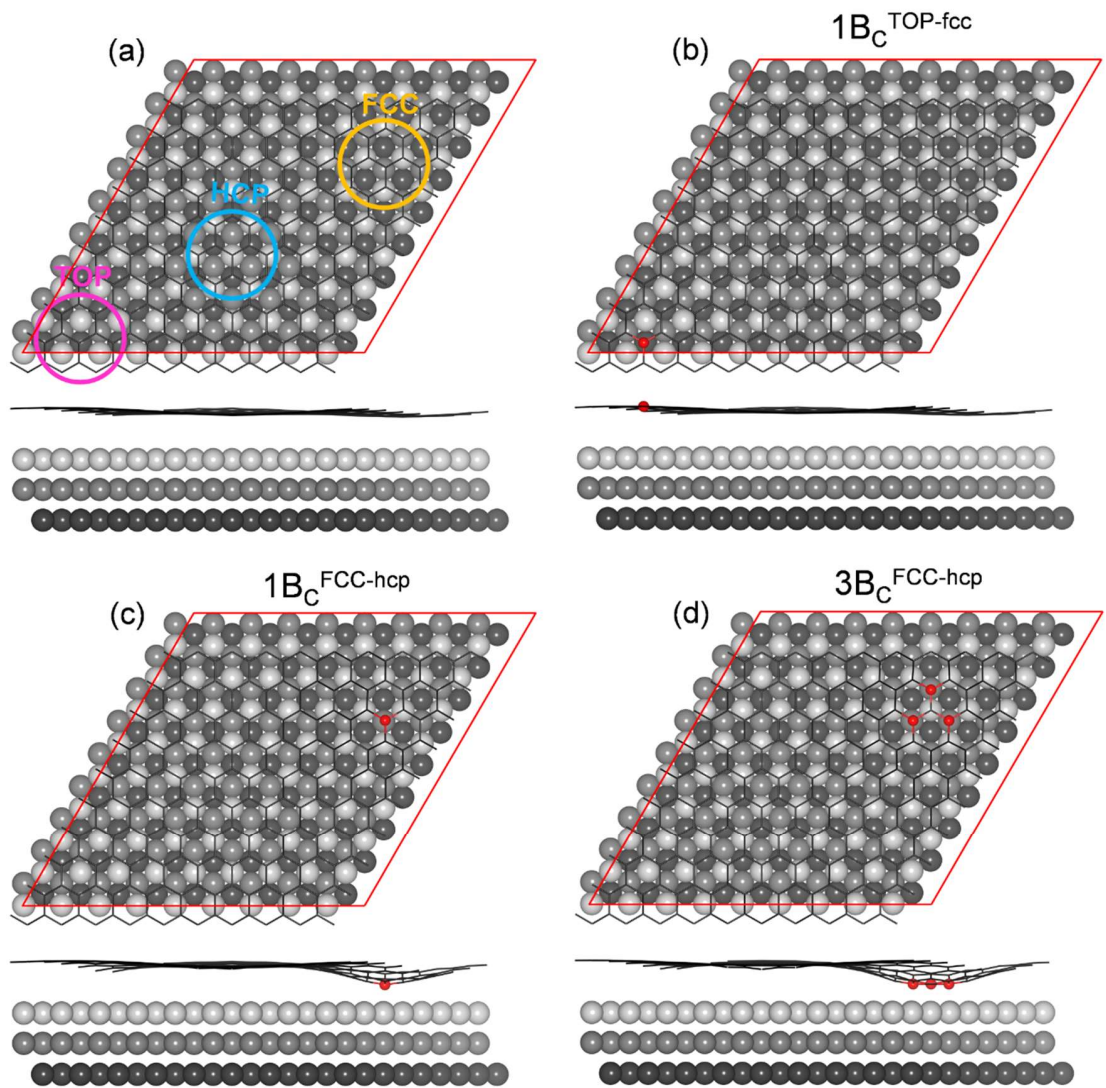

**Figure 3. Computed boron-doped graphene on Ir(111) structures.** Ball-and-stick models of (a) pristine and (b, c, d) boron-doped graphene supported on Ir(111) surface. Color coding: Carbon atoms of graphene are rendered in black (stick representation); substitutional boron atoms are rendered in red; iridium atoms of the first, second and third layer are rendered in light grey, grey and dark grey, respectively. (10×10) graphene on 9×9) Ir super-cell of the R0 domain depicted by the red parallelogram.

The variable shape of the "dents" observed in **Figure 1**a suggests that they might host multiple $B_C$ species, as proposed theoretically. In order to verify this experimentally, a closer inspection of the atomistic structure of the defects was performed by STM. The atomically-resolved image in **Figure 4**a shows the characteristic topographic appearance of the smallest occurring "dent" when measured at low bias voltage, consisting of a dark pseudo-triangular depression centered in one particular site of the graphene honeycomb network. This is attributed to a single $B_C$ based on the good agreement with the simulated STM image presented in **Figure 4**d, as well as the tendency of substitutional heteroatoms in graphene to exhibit 3-fold symmetric appearance [14]. Addition of other $B_C$ species leads to features with more extended and irregular geometries, as can be appreciated by comparing the experimental (**Figure 4**b,c) and simulated (**Figure 4**e,f) STM images. These findings further validate the assignment of the substitutional heteroatoms to boron and confirm their tendency to form small aggregates. Note that such aggregates generally do not include B-B bonds, as will be discussed below.

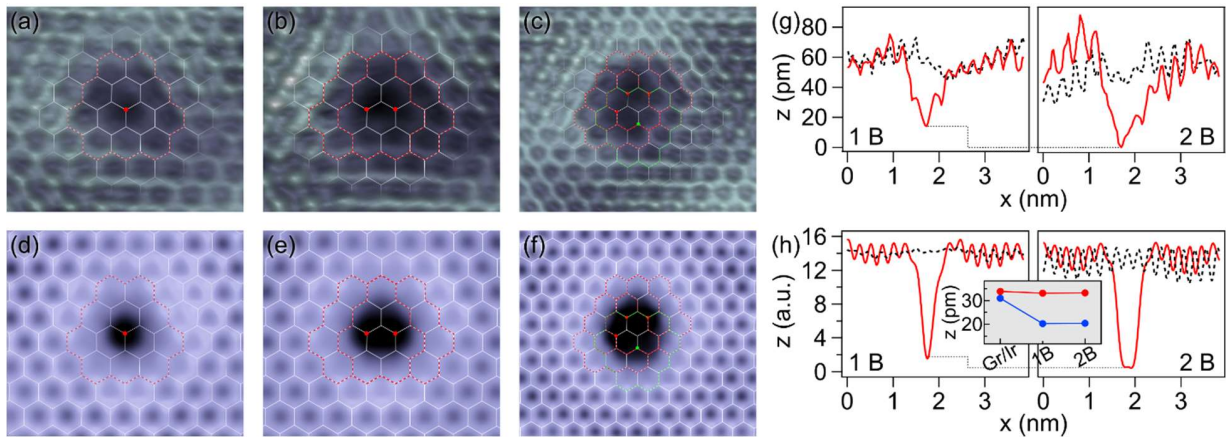

**Figure 4. Topographic appearance of boron substitutional heteroatoms in graphene on Ir(111).** (a-c) Experimental and (d-f) theoretical atomically-resolved STM images of (a, d) 1 $B_C$, (b, e) 2 $B_C$ and (c, f) 3 $B_C$ species occupying graphene low regions. (a-c) $V_{bias}$ = 0.01 V, and (d-f) $V_{bias}$ = 0.10 V with ILDOS value of $5 \times 10^{-5}$ $|e^-|/a_0^3$. Subtle 3D rendering was applied to images (a-c) for better visualization. Line profiles in (g) experimental and (h) theoretical STM images passing across $B_C$ (red lines) and far away from them (black dashed lines). Inset: Comparison between calculated average (red) and minimum (blue) graphene-iridium distance.

The fact that the substitutional boron species appear as depressions is surprising, as it contrasts the protrusion-like appearance of such defects reported for boron-doped graphene supported on numerous substrates studied to date [18,20,47,52,53]. Our DFT calculations suggest that this unique feature is mainly due to geometric bending of the graphene layer towards the Ir(111) substrate induced by the $B_C$ species, as detailed in **Figure 3**b and d. However, line profiles extracted from both experimental (**Figure 4**g) and simulated (**Figure 4**h) STM images, show that the depth of the depressions produced by 2 $B_C$ is larger than that of 1 $B_C$, while the calculated geometric bending of the graphene layer is about the same in the two cases (inset in **Figure 4**h). This points to electronic effects also contributing to the STM appearance of the single $B_C$ and aggregates.

Statistical analysis of the distribution of $B_C$ species was performed by counting the number of neighbors of a given $B_C$ as a function of their distance measured in number of atomic sites *n* of the graphene honeycomb lattice (see the inset in **Figure 5**a). For the analysis, multiple atomically-resolved STM images, like those shown in **Figure S4**, were used. The histogram in **Figure 5**a reveals that the $B_C$ species tend to occupy atomic sites within the same graphene sub-lattice, in agreement with the computed lower relative energies of $B_C$ pairs occupying the same sub-lattice over those in different sub-lattices, as shown **Figure 5**c. Interestingly, this behavior contrasts with the tendency to indistinctly occupy both sub-lattices predicted when boron atoms are incorporated in free-standing graphene, as reported in a previous computational study [54], and confirmed in this work. Very few $B_C$ species occupying adjacent sites were experimentally observed (*i.e.*, first neighbors), in agreement with a computed very high relative energy (+ 1.70 eV) with respect to two $B_C$ in the same sub-lattice, thus evidencing once more that formation of B-B bonds is highly unlikely. Equally important, the occurrence of neighbors (in both sub-lattices) decays with *n* (*i.e.*, the further apart the $B_C$ species are), hence explaining their tendency to distribute heterogeneously forming small aggregates as discussed above and shown in the STM images in **Figure 2** and **Figure 4**. The radius of the features arising from the $B_C$ aggregates is mostly limited to 8 honeycomb atomic sites (the occurrence of neighbors beyond this threshold becomes insignificant), which represents a maximum extension of the aggregates of around 7×7 graphene unit cells, slightly smaller than that of the moiré super-cell (between 9×9 and 10×10 depending on the rotational domain), hence indicating that the aggregates tend to be spatially constrained within a limited region of the 10×10 periodic unit. In addition, **Figure 5**b presents the occupancy distribution of $B_C$ species per aggregate, which peaks at 2 and decays until it vanishes around 6, having an expected value of 2.5 ± 1.9 $B_C$. Since the energy toll required to go from two isolated $B_C$ species (FCC-hcp + HCP-fcc) to aggregates of two (both in FCC-hcp) is small (~ 0.09 eV, see **Table 1**), it can be understood that the occupancy distribution is dictated by the number of available boron atoms during the growth.

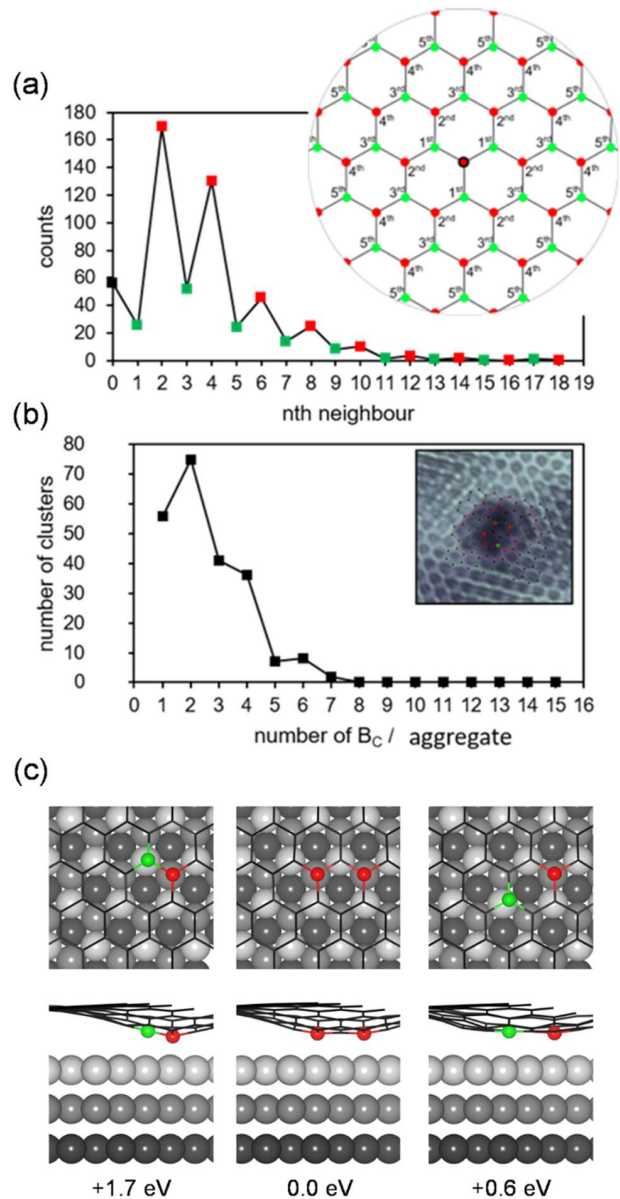

**Figure 5. Distribution of substitutional boron heteroatoms in graphene on Ir(111).** (a) Experimental radial distribution of $B_C$ species from a given $B_C$, measured in *n* atomic sites of the graphene honeycomb lattice. Red and green markers depict $B_C$ occupying the same or the opposite graphene sub-lattice with respect to the given $B_C$, as schematically shown in the inset. (b) $B_C$ occupancy distribution within aggregates. Statistics: 570 $B_C$ species having 1292 $B_C$ neighbors distributed within 225 aggregates. (c) Top and side view of 2 $B_C$ species occupying opposite (red/green balls) and the same (red/red balls) graphene sub-lattice on Ir(111).

Next, we will discuss how the electronic structure of graphene is affected by the incorporation of $B_C$ species, and by the local rippling that they produce. It is well known that substitutional boron in free-standing graphene induces p-type doping, due to its lower atomic number compared to carbon [55]. This results in an upshift of the Dirac cone with respect to the Fermi level, whose extent depends on the concentration of $B_C$ species: *i.e.*, the larger the amount of substitutional $B_C$ species in the layer, the more pronounced the p-type character of graphene. The expectation is that the presence of electronic holes in the π system of graphene will enhance its reactivity, altering its interaction with a metal substrate such as Ir(111).

In order to evaluate this effect more quantitatively, we calculated and compared the differential electron charge density maps following the deposition of pristine graphene and boron-doped graphene on Ir(111) (see 2D plots in **Figure 6**a and 3D ones in **Figure S4**). Starting with the un-doped system, we observe that graphene donates electrons to the metal forming a strongly interacting interface (see graphene/Ir(111) in **Figure 6**a), although the extent of such charge transfer varies within the moiré super-structure: *i.e.*, geometrically low regions (FCC and HCP) are transferring more electrons than geometrically high (TOP) regions, which are almost decoupled from the metal substrate. This scenario explains why graphene is slightly p-type doped when supported on Ir(111), as confirmed by calculated Bader charges (where graphene donates $\sim 1.6 \times 10^{-3}$ e$^-$/carbon atom) and projected (PDOS) density of states (**Figure 6**b). It now becomes interesting to investigate what happens when $B_C$ species are present. The bottom panel of **Figure 6**a shows how the substitutional boron strongly perturbs the graphene/Ir(111) interaction by inducing a larger charge transfer from the underlying Ir atoms to the carbon atoms directly coordinated to the boron atoms (see charge accumulation in **Figure 6**a) and facilitating the local down-bending of graphene. It is worth noting that the perturbation induced by $B_C$ is quite local, since, moving further away from the defect, the charge plot returns to be similar to that of graphene/Ir(111).

In order to verify the doping scenario proposed theoretically, local and non-local measurements of the electronic structure of the boron-doped graphene were conducted by STS and ARPES, respectively. In **Figure 6**c, a series of d$I$/d$V$ curves measured by STS on individual aggregates containing one or more $B_C$ species and on un-doped regions is presented. It shows that the boron-doped graphene remains slightly p-doped, as the minimum of intensity, identified as the Dirac point, lies 24 ± 5 mV above the Fermi level regardless of the region. Note that the local maximum at ~ -180 mV is attributed to the Ir(111) surface state [56], slightly up-shifted due to the presence of graphene [57]. The fact that neither the Dirac point position nor the shape of the d$I$/d$V$ curves vary between the un-doped and doped regions, in agreement with the calculated PDOS (**Figure 6**b), reveals that the local electronic structure near the Fermi level remains largely unaltered upon introduction of $B_C$ species. Analogous conclusions can be reached based on the analysis of the computed workfunctions (Table S1) since we observe that the value for 3B-doped FCC graphene/Ir ($3B_c^{FCC-hcp}$) is essentially the same as for graphene/Ir (4.63 vs 4.67 eV, respectively).

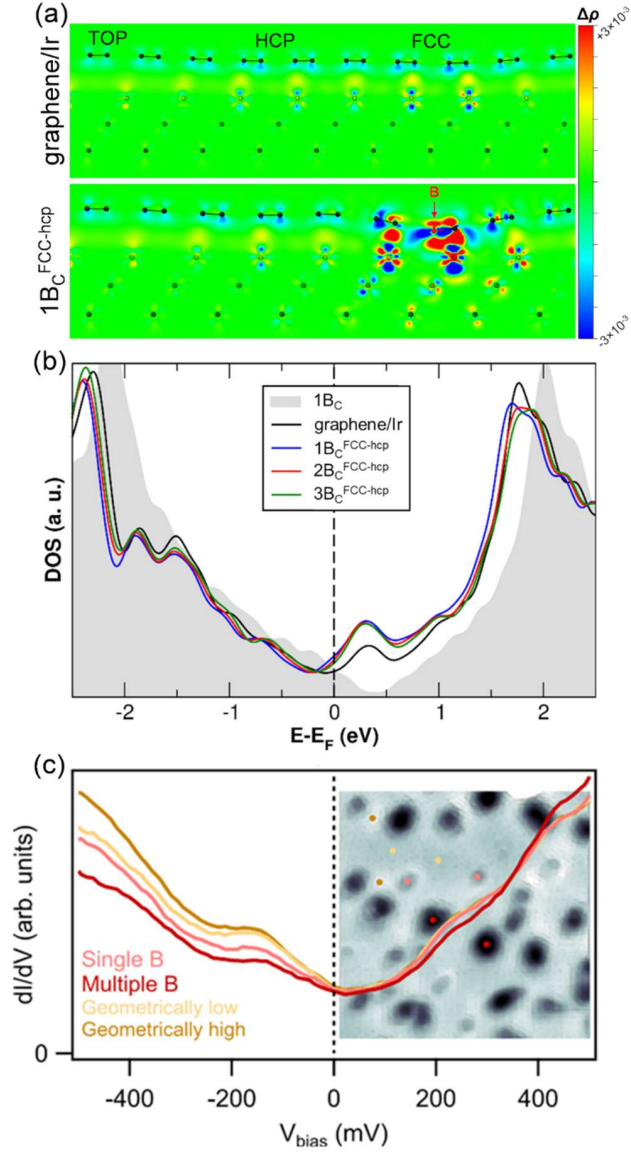

**Figure 6. Local electronic structure of boron-doped graphene on Ir(111).** (a) 2D charge density difference plots (corresponding 3D plots are shown in **Figure S5**) for pristine graphene and 1 $B_C^{FCC-hcp}$ supported on Ir(111) surface. The color scale indicates electron depletion and accumulation ranging from blue to yellow/red, respectively. Isosurface level is 0.001 e⁻/bohr³ for both systems. Geometries have been superimposed to aid interpretation. (b) PDOS on graphene for pristine and boron-doped graphene supported on Ir(111). The Fermi level is scaled to zero and indicated by a blacked dashed line. (c) d$I$/d$V$ spectra acquired on the different regions of a boron-doped graphene domain (R18) on Ir(111) shown in STM image of the inset ($V_{bias}$ = 0.60 V, 11 nm lateral size, gentle 3D render has been applied for better visualization). Stabilization conditions: $V_{bias}$ = 0.50 V and $It$ = 0.5 nA, lock-in modulation voltage $V$ = 5 mV.

In addition, ARPES measurements of boron-doped and un-doped layers allow to assess the impact of the $B_C$ species on the electronic band structure of graphene (see **Figure S1** for more details about the un-doped reference sample). In the spectra taken along the Γ→K direction of graphene's (R0) Brillouin zone [58], a prominent graphene π band appearing at ~ 8 eV that crosses the Fermi level near $K_{graphene}$ is identified in both doped and un-doped graphene (**Figure 7**a, b), alongside with other states of attenuated intensity attributed to the underlying Ir(111) [57,59]. Note that the high-momentum branch of the Dirac cones is hardly visible due to interference between photoelectrons emitted from the two sub-lattices of graphene [60], and that the increased background intensity of the boron-doped spectra is attributed to the presence of a manifold of rotational domains (see **Figure 2**d, e and **Figure S4**). A comparison between the linearly dispersing π band of un-doped and boron-doped graphene near the Fermi level (*i.e.*, the Dirac cone) is presented in **Figure 7**c, in which the right panel has been mirrored along $k_x$ to ease comparability. No appreciable change in the dispersion of the Dirac cone nor in the energy position of the Dirac point (from 0.24 ± 0.15 to 0.27 ± 0.15 eV, see **Figure S6**) is observed, in stark contrast to the binding energy shift (up to ~ 1 eV) expected for boron-doped free-standing graphene at low boron concentrations (< 5 %) [46]. Hence, the fact that the Dirac cone remains unperturbed upon the introduction of substitutional boron atoms confirms their negligible doping effect on Ir-supported graphene suggested by theoretical calculations.

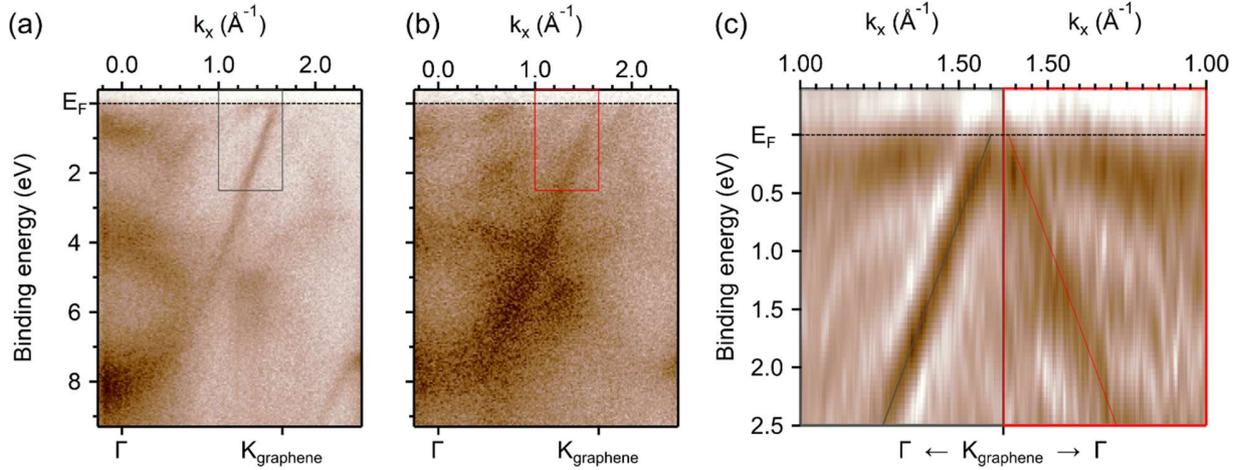

**Figure 7. Occupied electronic structure of boron-doped graphene on Ir(111).** ARPES spectra of (a) un-doped graphene/Ir(111) and (b) boron-doped graphene/Ir(111) measured along the Γ→K direction. (c) Second derivative plot of the magnified regions marked by the colored rectangles in (a) and (b). Grey and red lines are a guide to the eyes that denote the position of the Dirac cone near the Fermi level for un-doped and boron-doped graphene, respectively. Spectra were measured at a 22.5° tilt angle from normal emission in order to visualize the first and part of the second Brillouin zone of graphene.

Finally, we venture to rationalize why the presented method yields formation of ordered arrays of $B_C$ species, in striking contrast to others reported earlier. First, the use of a highly corrugated moiré superlattice, like that induced by Ir(111) or Ru(0001) [27,47], is instrumental to enable site-selective pinning of the $B_C$, as deduced from our DFT calculations. Second, incorporation of the heteroatoms during graphene growth, promoted here by the carbon and boron-containing molecular precursor, could play an important role because of an enhancement of $B_C$ segregation at elevated temperatures, facilitating the access to the favorable moiré regions. A similar scenario was demonstrated by Zabet-Khosousi *et al.,* [61] by comparing the incorporation of nitrogen during and after graphene growth, and cleverly used by Camilli *et al.* [62] to engineer arrays of 0D graphene dots embedded within a 2D BNC alloy.

## 4. Conclusions

In conclusion, we have shown how the incorporation of boron in graphene during its growth process on Ir(111) results in 2D arrays of boron species structurally embedded in graphene. Combining atomic-scale and surface-averaged characterization complemented by computational modelling reveals that this structure is intimately related to the strong corrugation of the moiré superlattice of graphene on Ir(111), as the formation of boron substitutional species is energetically favored only within its geometrically low regions. The impact of the boron species on graphene's doping level is found to be marginal, as the holes that they introduce in graphene are compensated by the uptake of electron charges from the substrate through a stronger interfacial interaction. This process is facilitated by a local bending of the graphene layer towards the substrate that induces charge accumulation around the heteroatoms. As a result, the moiré-modulated electron density and rippling of graphene become greatly enhanced, making this material highly appealing for applications in the fields of metal-free catalysis, gas sensing, or nano-electronics.

ASSOCIATED CONTENT

**Supporting material**

Figures S1 – S6: Un-doped graphene on Ir(111) reference sample (XPS and LEED); schematic of the calculated configurations of $B_C$ in graphene/Ir(111) and corresponding CLS calculations; schematic of the calculated $B_C$ positions leading to the relative energies reported in **Table 1**; STM characterization of diverse rotational domains of boron-doped graphene on Ir(111); 3D charge density difference plots; extended view of the ARPES data presented in **Figure 7**c. Table S1: Calculated workfunctions for free-standing and Ir-supported pristine and B-doped graphene. Table S2: XPS fitting parameters used to model the XPS core levels shown in **Figure** 1 and **Figure** S1.

**Corresponding Authors**

Marc G. Cuxart – Physics Department E20, Technical University of Munich, James-Franck-Str. 1, 85748 Garching, Germany; orcid.org/0000-0002-9085-1225; Email: marc.gonzalez-cuxart@tum.de

Cristiana Di Valentin – Dipartimento di Scienza dei Materiali, Università di Milano Bicocca, via R. Cozzi 55, 20125 Milano, Italy; orcid.org/0000-0003-4163-8062; Email: cristiana.divalentin@unimib.it

Willi Auwärter – Physics Department E20, Technical University of Munich, James-Franck-Str. 1, 85748 Garching, Germany; orcid.org/0000-0001-9452-4662; Email: wau@tum.de

**Authors**

Daniele Perilli – Dipartimento di Scienza dei Materiali, Università di Milano Bicocca, via R. Cozzi 55, 20125 Milano, Italy; orcid.org/0000-0002-3082-3986

Sena Tömeke – Physics Department E20, Technical University of Munich, James-Franck-Str. 1, 85748 Garching, Germany.

Joel Deyerling – Physics Department E20, Technical University of Munich, James-Franck-Str. 1, 85748 Garching, Germany.

Felix Haag – Physics Department E20, Technical University of Munich, James-Franck-Str. 1, 85748 Garching, Germany;

Matthias Muntwiler – Paul Scherrer Institut, Villigen 5232, Switzerland; orcid.org/0000-0002-6628-3977

Francesco Allegretti – Physics Department E20, Technical University of Munich, James-Franck-Str. 1, 85748 Garching, Germany. orcid.org/0000-0001-6141-7166.

**Author Contributions**

‡M.G.C and D.P. contributed equally to this work.

**Notes**

The authors declare no competing financial interest.

**Acknowledgments**

The authors thank H. Sachdev for suggesting the use of the BTHF precursor. M.G.C. acknowledges funding from the European Union's Horizon 2020 Research and Innovation Programme under the Marie Skłodowska-Curie grant agreement no. 892725 (WHITEMAG project). S.T. and W.A. acknowledge the MSCA-ITN-ETN STiBNite (no. 956923). The Paul Scherrer Institut, Villigen, Switzerland is acknowledged for providing synchrotron radiation beamtime at the PEARL beamline of the SLS. C.D.V. acknowledges funding from MIUR through the project "MADAM: Metal Activated 2D cArbon-based platforMs" PRIN2017 grant No. 2017NYPHN8 and from European Research Council (ERC) under the European Union's HORIZON2020 research and innovation programme (ERC Grant Agreement No [647020]).